\documentstyle[12pt,epsf]{article}

\begin{document}
\title
{PDF's Of The Burgers Equation On The Semiline With
Fluctuating Flux At The Origin}
\author
{Jahanshah Davoudi {\footnote{\it jahan@theory.ipm.ac.ir}}
and 
Shahin Rouhani \\
\\
\it
Dept. of Physics , Sharif University of Technology,\\
\it P.O.Box 11365-9161, Tehran, Iran.\\
\\ \it  Institute for Studies in Theoretical Physics and 
Mathematics
\\ \it Tehran P.O.Box: 19395-5746, Iran. 
}
\maketitle
\begin{abstract}
We derive the asymptotic behaviour of the one point probability density for
the inhomogeneous shock slopes in the turbulent regime, when a Gaussian
fluctuating flux at origin derives the system.
We also calculate the time dependence of the $x_{f}$ beyond which there won't exist any
velocity shocks as $x_{f}\cong t^{3/4}(\log{t})^{1/4}$. 
We argue that the stationary state of the problem would be equivalent with the
long time limit of the diffusion equation with a random source at origin.
\end{abstract}

\def\be{\begin{equation}}
\def\ee{\end{equation}}
\def\bear{\begin{eqnarray}}
\def\eear{\end{eqnarray}}
\newpage

\section{Introduction}
Burgers equation \cite{b1} initially was suggested as a model for describing the
propagation of nonlinear waves in dissipative media \cite{b2}. Burgers equation
with a stochastic source has been suggested as a phenomenological equation covering
the behaviour of a large universality class of nonequilibrium growth models
\cite{k}.Under a nonlinear transformation it can also be transformed into a model
for diffusion in Random landscapes which is itself related to a rich class of
phenomena like, Directed Polymers in Random media \cite{bhz}, wetting \cite{fda},
Random Magnets \cite{hen} and Pinning of vortex lines in superconductors \cite{gb}.
This equation is thoroughly studied as an approximate model describing the 
formation of large scale structures in Universe \cite{zel}.
Burgers Equation with a noise term has also been studied as a toy model for 
fully-developed turbulence with using a variety of methods \cite{fdt}. The full 
description of this equation is still far from clear but the introduction
of new methods has given a better understanding of the relationship
between different realms of physics. The large time solutions of 
deterministic Burgers equation with random initial conditions has been
investigated by the turbulence community for mimicking the behaviour of 
decaying turbulence \cite{dt}. 
An alternative way of introducing fluctuations is to consider  
random initial conditions. The physical relevance for instance comes from 
considering a variable injection of fluid into a semi-infinite slab
of soil.

Semiline solutions of Burgers equation with a given deterministic
flux at origin has been
derived \cite{lill}, 
which consistently reproduces the inhomogeneous shocks of Burgers
equation at large spatial scales. By allowing random flux
, we observe that shocks are present at large times but with random
values and the velocities become ordered.

\section{The Semiline Solution}

Consider the Burgers equationon the semiline, with the initial and boundary
condition:
\be
\label{1}
u_t + uu_x =\nu u_{xx}\nonumber  
\ee
\bear
u(x,0) = 0 \ \ \ 0\leq x\leq\infty  
\eear
\be
\frac{u^2(0,t)}{2} + \nu u_x(0,t) = g(t)
\ee

The above initial value problem is solved through a generalised Hopf-Cole transformation
\cite{lill}.
\be
u(x,t)=\frac{2}{\sqrt{\pi}}\frac
{\int_0^t dt'\frac{g(t')c(0)}{\sqrt{4\nu(t-t')}}exp(-\frac{1}{\nu} (\frac{x^2}{4(t-t')}-\int_0^{t'}g(t'')dt'') )}
{c(t)-\frac{1}{\nu}\int_0^tdt'g(t')exp(\frac{1}{\nu}\int_0^{t'}g(t'')dt'') erf(\frac{x}{ \sqrt{4\nu(t-t')}} )}\label{uxt},
\ee
where:
\be
c(t)=c(0) e^{\int_0^{t}g(t')dt'}
\ee
In the limit where $\nu$ tends to zero we can expand equation(\ref{uxt}) for regions 
where $\frac{x}{\sqrt{4\nu t}}\gg1$:
\be
u(x,t)\cong\frac{2}{\sqrt{\pi}}\frac
{\int_0^t dt'\frac{g(t')c(0)}{\sqrt{4\nu(t-t')}}exp(-\frac{1}{\nu} (\frac{x^2}{4(t-t')}-\int_0^{t'}g(t'')dt'') )}
{1+\frac{1}{\nu}\int_0^tdt'g(t')(\frac{\sqrt{4\nu(t-t')}}{\sqrt{\pi}x})exp(-\frac{1}{\nu} (\frac{x^2}{4(t-t')}-\int_0^{t'}g(t'')dt'') ) }\label{tauxt},
\ee
Following Burgers\cite{b2}, we note that the main contribution to the integral in
(\ref{tauxt}) come from points which maximize the exponent:
\be
h(x,t,t')=-\frac{x^2}{4(t-t')} + \int_0^{t'}g(t'')dt''
\ee
After some manipulations, equation (\ref{tauxt}) reduces to
\be
u(x,t)\cong\frac{x}{t-t_i}
\ee
where $t_i(x,t)$ are the absolute maxima of $h(x,t,t')$. These are indeed the
shock waves discovered by Burgers through an elegant geometric construction.

Burgers geometric construction finds the maxima of $h(x,t,t')$ by defining two
functions:
\bear
S(t')&=&\frac{x^2}{4(t-t')}+H \label{7},\\
G(t')&=&\int_0^{t'}g(t)dt \label{8},
\eear
where $H$ is a constant.
For a given $t$, one looks for the maximum value of $H$, called $H^{*}(t_i)$, where
the hyperbola $S(t')$ just touches the curve $G(t')$.
For large values of $x$ the hyperbola $S(t')$ flattens, and it may touch $G(t)$
at two points. When this happens we have a shock wave, thus the solutions are 
comprised of linear parts, $\frac{x}{t-t'}$ . In small regions of space, intervening
the linear parts, the velocity profile is more complicated. The long time velocity
profile is a series of shocks with reducing slopes. This can be understood
as response of the system to impulses imported to it at origin, at consecutive 
times.

\section{One Point Probability Density}

Let us now analyse the same problem for the case of the random impulse
at origin and assume that $g(t)$ is a random variable with Gaussian distribution:
\bear
\langle g(t)\rangle&=&0\\
\langle g(t)g(t')\rangle&=&2{\sigma_{g}}^2\Delta(t-t')
\eear
Where $\Delta$ is any reasonable correlation which decays with growing argument.
This means that $G(t)$, as defined in equation(\ref{8}) is a stochastic process and
we assume that the variance of this process depends linearly on time so that it
a nonhomogeneous process with some correlation time $\tau$ overwhich its values
are correlated and beyond that uncorrelated.
Giving address to the geometrical construcion discussed in the previous section
the problem reduces to finding the maxima of a nonhomogeneous process $G(t)$.
For this purpose we can adopt the methods previously used for the random
initial conditions\cite{dt}.
Consider the commulant probability of the event that the absolute maximum 
of $h(x,t,t')$ occurs in the interval $[t'_1,t'_2]$,that is,
\be
F(H,[t'_1,t'_2])=Prob(\frac{x^2}{4(t-t'_i)} + G(t'_i) < H ; t'_i\in[t'_1,t'_2])
\ee
If the correlation time of $G(t)$ is smaller than the mean difference of two adjacent
$t'_{i}$ 's:
\be
\overline{(t'_{i+1}-t'_{i})}>\frac{1}{{\sigma_{g}}^2}
\ee
then one can assume independence for regions inside the interval $[t_1,t_2]$
and outside it. Thus the probability of finding $t_i\in[t'_1,t'_2]$ is :
\be
Prob(t'\in[t'_1,t'_2])=\int_{-\infty}^{\infty}F(H,[0,t]-[t'_1,t'_2])P(H,[0,t]-[t'_1,t'_2])dH\label{20}
\ee

where \ \ \ \ \ $P=\frac{dF}{dH}$.\vspace{2mm}

Indeed in the Poissonian approximation we have :

\be
F(H,[t_1,t_2])\cong e^{-N(H,[t_1,t_2])}\label{21}
\ee
where $N$ is the expected number of crossings between $G(t')$ and $H+\frac{x^2}{4(t-t')}$ in
the interval $[t_1,t_2]$
and is given 
\be
N(H,[t_1,t_2])=\langle\int\delta(\Phi)\{\frac{\partial{\Phi}}{\partial{t}}\}^{+}dt'\rangle\label{22}
\ee
where\ \ \ $\Phi= \frac{x^2}{4(t-t')} - G(t')$ and $\{x\}^{+}=x \theta(x)$.
The average is taken over the joint distribution of $g(t)$ and $G(t)$, and
there is the constraint that at the point of intersection $G(t)$ has the positive
slope.
After some manipulations and in the limit
when $\frac{x}{\sqrt{\sigma_g}}\ll t$ and $H\gg\sigma_{g}t$ we get:
\be
N(H,[t_1,t_2])\cong\int_{t_1}^{t_2}dt'\frac{\sqrt{{\sigma_g}^2 t'}}{H\sqrt{2\pi}} e^{-\frac{{ (H^2+\frac{x^2}{4(t-t')}) }^2     }     {2t'{\sigma_g}^2 }}\label{24}
\ee
and consequently $F(H,[0,t])$ will become:
\be
F(H,[0,t])\cong\ e^{-\int_{0}^{t} dt'\frac {\sqrt{{\sigma_g}^2t'}}{S(t')\sqrt{2\pi}} e^{-\frac{{S(t')}^2 }{2t'{\sigma_g}^2} }}
\ee
where \ \ \ \ $S(t')=H+\frac{x^2}{4(t-t')}$.
For calculating the exponent we use from the saddle point approximation in the integral of
the exponent. So in the large time limit we simply have the following saddle point
solution for $t^{*}$ :
\be
t-t^{*}\cong\ \frac{x^2}{4H}+\sqrt{\frac{tx^2}{4H}}
\ee
Before proceeding further we wish to find an estimate for $x_f$ where the shock
would disappear. Taking a closer look at the $t^{*}$ we find that when $N(H,[0,t])$ is maximum around $t^{*}$
the cummulant probability has it's most probable value and the contact point $t^{*}$
would be the most probable contact point. So a good estimate for $x_f$ would be
found by setting the $t^{*}\cong 0$,from which we get
\be
x_f\cong\sqrt{4tH}\label{xf}
\ee

We believe that the main contribution for the integral(\ref{20}) would be around
a characteristic value of $H$,say $H^{*}$ such that the cummulative probability $F(H,[0,t])$
gives its dominant contribution. On the other hand we find $H^{*}$ from
\be  
N(H,[0,t])\cong 1\label{ex1}
\ee
or
\be
\frac{\sqrt{{\sigma_g}^2t'}}{S(t')\sqrt{2\pi}}e^{-\frac{{S(t')}^2}{2t'{\sigma_g}^2}}\cong1
\ee
Intuitively this characteristic value gives the border beyond which there would be
just one intersection between the process $G(t')$ and the curve $S(t')$. 
After doing some algebra we find $H^{*}$ in the large time limit and in the domain
$x\ll{x_{f}}$ as
\be
H^{*}\cong\sigma_{g}\sqrt{t\log{t}}\label{hs}
\ee
By substituting the derived $H^{*}$ from (\ref{hs}) in expression (\ref{xf}) we
find the time dependence of $x_{f}$
\be
{x_{f}}^2\cong\sigma_{g}{t}^{3/2}{(\log{t})}^{1/2}
\ee
When $t\gg\sqrt{\frac{x^2 t}{4H}}$ or $x\ll{x_{f}}$ the cummulative probability density
will be:
\be
F(H,[0,t])\cong e^{ -\frac{\sigma_{g}\sqrt{t}}{\sqrt{2\pi}(H+\frac{x\sqrt{H}}{\sqrt{4t}})} e^{-\frac{{(H+\frac{x\sqrt{H}}{\sqrt{4t}}) }^2 }{2{\sigma_{g}}^2 t}}}
\ee
The asymptotic form of $F(H,[0,t])$ may be computed for large values of $H=H^{*}+\beta z$
and the result comes out to be the well known Gumbel distribution\cite{gum}.
\be
F(H,[0,t])\cong e^{-e^{-z}}
\ee
where
\be
\beta=\frac{{\sigma_{g}}^2t}{H^{*}+\frac{x\sqrt{H^{*}}}{\sqrt{4t}}}\label{beta}
\ee
By substituting(\ref{21}) in (\ref{20}) an taking an integration by part
we will get
\be
P(t'\in[t',t'+\Delta t']))=\int_{0}^{\infty}N(H,[t_1,t_1+\Delta t])\frac{dF(H,[0,t])}{dH}dH\label{28}
\ee
We have calculated the probability distribution in (\ref{28}) by approximating $N(H,[t_1,t_1+\Delta t])$
in (\ref{24}) for deviations from $H^{*}$ and substituting in (\ref{28}):
\be
P(t'\in [t',t'+ \Delta t'] | x,t)= \gamma \Gamma(\alpha\beta+1,1)\label{prob1}
\ee
Where $\alpha$ and $\gamma$ are defined as
\bear
\gamma&=&\frac{\sqrt{2{\sigma_{g}}^2t'}}{H^{*}}e^{\frac{1}{2{\sigma_{g}}^2t'} {(H^{*}+\frac{x^2}{4(t-t')}) }^2}\label{gam}\\
\alpha&=&\frac{S^{*}(t')}{{\sigma_{g}}^2 t'}\label{alph}
\eear
We have the following intuitive picture that as one measures the slope in a linear part of the velocity profile in a fixed
spatial point the most probable value of is a decreasing function of $x$.
By computing the numerical integration of the probability density one can easily see that
for $x's$ greater $x_{f}$ which is given by (\ref{xf}) the total probability of occuring
a contact is vanishing.
By changing variable from $t'$ to $u(x,t)$ ,
one can easily find the probability density of $u(x,t)$
for a fixed $x$ and $t$ as:
\be
P(u|x,t)=\frac{x}{u^2 H^{*}}\sqrt{2{\sigma_{g}}^2(t-\frac{x}{u})}e^{-\frac{1}{2{\sigma_{g}}^2(t-\frac{x}{u})}{(H^{*}+\frac{xu}{4})}^2} \Gamma(\frac{(H^{*}+\frac{xu}{4})t}{H^{*}(t-\frac{x}{u})}+1,1 )
\ee
We have numerically integrated the average shock's slope and velocity for different
values of $x$ in a fixed observation time, and have plotted their character in
different graphs (figures 1,2).

\vspace{0.3cm}
\section{Two Point Probability Density}

In this section we generalise the above idea for calculating the two point statistics
of the shocks. By using the geometrical picture of the contacts between the
hyperbola and the random process $G(t')$, every information about the two point statistics
can be mapped to the different cathegories of geometrical events.
If we are interested in the two point statistics in two different times
$t_1,t_2$ but at the same fixed point $x$, the correspondent event would be the
contact of two parabolas with different vertical asymptotic axes at $t_1,t_2$ 
, with the random process $G(t')$ in two points say $t'_1,t'_2$ without ever intersecting the
other parts of it. Similarly the two point statistics in two different spatial points
$x_1,x_2$,and at a fixed observation time $t$, would be translated to definite geometrical events. 
Technically the same structure for calculating these quantities can be applied in principle.
For example in calculating the two point probability density $P(u_1,u_2)$ on which
$u_1=u(x_1,t)$ and $u_2=u(x_2,t)$, we use from the geometrical construction
as in (fig 4). According to the figure, there are two situations distinguished
which are in correspondence to existence and nonexistence of shocks between the two contact points 
$t'_1$ and $t'_2$. So we proceed to find    
\be 
P(t'_1 \in L_1,t'_2 \in L_2) = P_{p}(t'_1,t'_2) + \delta(t'_1 -t'_2) P_{a}(t'_1,t'_2)\label{pp}
\ee
where $P_p$ corresponds to the peresence and $P_{a}$ corresponds to absence of shocks
between two points $t'_{1}$ and $t'_{2}$.
Existence of Dirac delta function in (\ref{pp}) is related to the fact that in large spatial
scales $x\in [x_1,x_2]$ and in the absence of discontinuties between $t'_1$ and $t'_2$  
the two hyperbolas should approximately contact near one of the local maxima 
of $G(t)$, thus $t'_1 \cong t'_2$.

By reffering to (fig 3) and adapting the techniques 
of previous section for calculating $P_p(t'_1,t'_2)$ we write:
\bear
&&P_{p}(t'_1,t'_2)=\int dH_1 \int dH_2 e^{-N(H_2,[0,t'_1])} e^{-N(H_2,[t'_1+\Delta t,t^{*}]) } e^{-N(H_1,[t^{*},t'_2])}\times\nonumber\\
&&\times e^{-N(H_1,[t'_2+\Delta t,t]) }\frac{d}{dH_1}e^{-N(H_1,[t'_1,t'_1+\Delta t]) }\frac{d}{dH_2}e^{-N(H_2,[t'_2,t'_2+\Delta t]) }\label{p12}
\eear
on which $t^{*}$ is the intersection point of two hyperbolas, that is:
\be
t-t^{*}=\frac{x_1^2-x_2^2}{4(H_2-H_1)}
\ee
It is clear that in the limit $\sigma_g t \gg H^{*}$ and for large times, the main contribution
in (\ref{p12}) is related to region near $H^{*}$ given in (\ref{hs}). Thus by  
defining $H_1=H^{*} +\beta z_1$ and $H_2=H^{*} +\beta z_2$ 
the integration of (\ref{p12}) will give the two point statistics .
The main point in calculating the integral is the existence of the same lavel $H^{*}$
over which the extremes of the proposed process $G(t)$ have their major contribution.
Calculations in this direction will be pursued in future.

\vspace{0.3cm}             
\section{Diffusion Limit}

One of the interesting questions is related to stationary state of the system
and its relationship to our picture. We believe that the stationary regime in
our problem is governed by the diffusion equation stirred at origin.
For showing the validity of our statement we should compare the average of the
two terms in the denominator of the expression for $u(x,t)$ given in (\ref{uxt}).
Before taking an integration by parts the denominator can be written as
\be
c(0)erf(\frac{x}{\sqrt{4\nu t}})+(\frac{x}{\sqrt{\pi}})\int_{0}^{t}dt'\frac{c(t')}{\sqrt{4\nu{(t-t')}^3}}e^{-\frac{x^2}{4\nu(t-t')}}\label{den}
\ee
After taking the average over the distribution of $g(t)$ one can easily find
that the average of $c(t)$ is
\be
\langle c(t)\rangle =c(0)e^{\frac{{\sigma_g}^2t}{2{\nu}^2}}\label{av_c}
\ee
By substituting from (\ref{av_c}) into the (\ref{den}) and changing the variable
of integration from $t'$ to $w$ we will end up with
\be
c(0)[erf(\frac{x}{\sqrt{4\nu t}})+\frac{x}{\sqrt{4\nu}}e^{\frac{{\sigma_g}^2t}{2{\nu}^2}}\int_{\frac{1}{t}}^{+\infty}dw{w}^{-\frac{1}{2}}e^{-(\frac{x^2}{4\nu})w-(\frac{{\sigma_g}^2 }{2{\nu}^2}))(\frac{1}{w})}]
\ee
Because we are interested in steady state,we should find the large time limit
of the denominator. The limit of the integral can be written in terms of the modified
Bessel function $K_{\frac{1}{2}}[\frac{x\sigma_g}{\sqrt{2{\nu}^3}}]$, so the
denominator in large time limit can be written as
\be
c(0)e^{\frac{{\sigma_g}^2t}{2{\nu}^2}(1-\frac{x}{\sqrt{\nu t^{*}}})}
\ee
where
\be
t^{*}=\frac{{\sigma_g}^2{t}^2}{2{\nu}^2}
\ee
It is obvious that in the limit where $t\rightarrow+\infty$ the denominator behaves as
(\ref{av_c}) which means that the integral in the denominator of (\ref{uxt})
would be negligible in average as if $u(x,t)$ satisfies the diffusion equation
with random source at origin.

\vspace{0.2cm}             
\section{Conclusion}
We have found the one point statistics of the Burgers equation with random boundary
condition on a semiline. The random boundary condition in the problem can be translated as a
random momentum input so the total momentum is not conserved. We have solved the
problem by a geometrical construction which relates the original problem to the
Extreme universality of the input noise in time(\cite{gum,bm}). In the case of a non-homogeneous
noise with a variance proportional with $t$ we found the Gumbel universality
class(\cite{gum}). The overall picture which we have found is that in a snapshot of the velocity
profile there is a moving point $x_{diff}\cong\sqrt{\nu t^{*}}$ which gives a crossover region from the diffusive
behaviour to shock structure such that with the passage of the time the diffusive region 
will grow and in the infinite time limit the diffusive regime would be dominant in the whole semiline.\\
Recently it has been shown that there are some evidences which
 relates different Extreme universalities with different Replica
symmetry breaking schemes(\cite{bm,mm}) and specifically the relation between the
Gumbel universality with to the
one step Replica symmetry breaking is checked in some models. Checking another
paradigm we think that it would
be important to apply the Replica method on this problem too.
Another problem worthy of investigation is the calculation of intermittency exponent
structure function, and checking whether the nonhomogenity alters the intermittency
exponent.
Finally we think that the KPZ and directed polymer analogs of our problem which have
been studied recently(\cite{ml}) is relevant too because the known results in those
realms are studied under the homogeneous noise and the presence of a nonhomogeneous
noise might change the universality classes in those problems.
\\
\\
{\bf Acknowledgement:}
\\
J. Davoudi would like to thank M.R. Ejtehadi, B. Davoudi and R. Rahimitabar
for useful discussions.

\newpage
{\large Figure Captions}

Figure 1- The behaviour of average shock slope in terms of $x$.
          $\sigma_{g}=1$ and the observation

           time is $t=1000$. The best fit for the curve is $\langle\lambda\rangle\cong\frac{1}{a+bx^{3/2}}$
            where $a\cong 182$ and $b\cong 0.05$.

Figure 2- The average velocity in terms of $x$.$\sigma_{g}=1$
          and the observation time is $t=1000$. The best fit for the curve is $\ln{\langle u(x) \rangle}\cong a+b\frac{\ln{x}}{x}$
          where $a\cong-0.1$ and $b\cong-16$.

\newpage
\begin{figure}
\begin{center}
\epsfxsize=10cm
\epsfbox{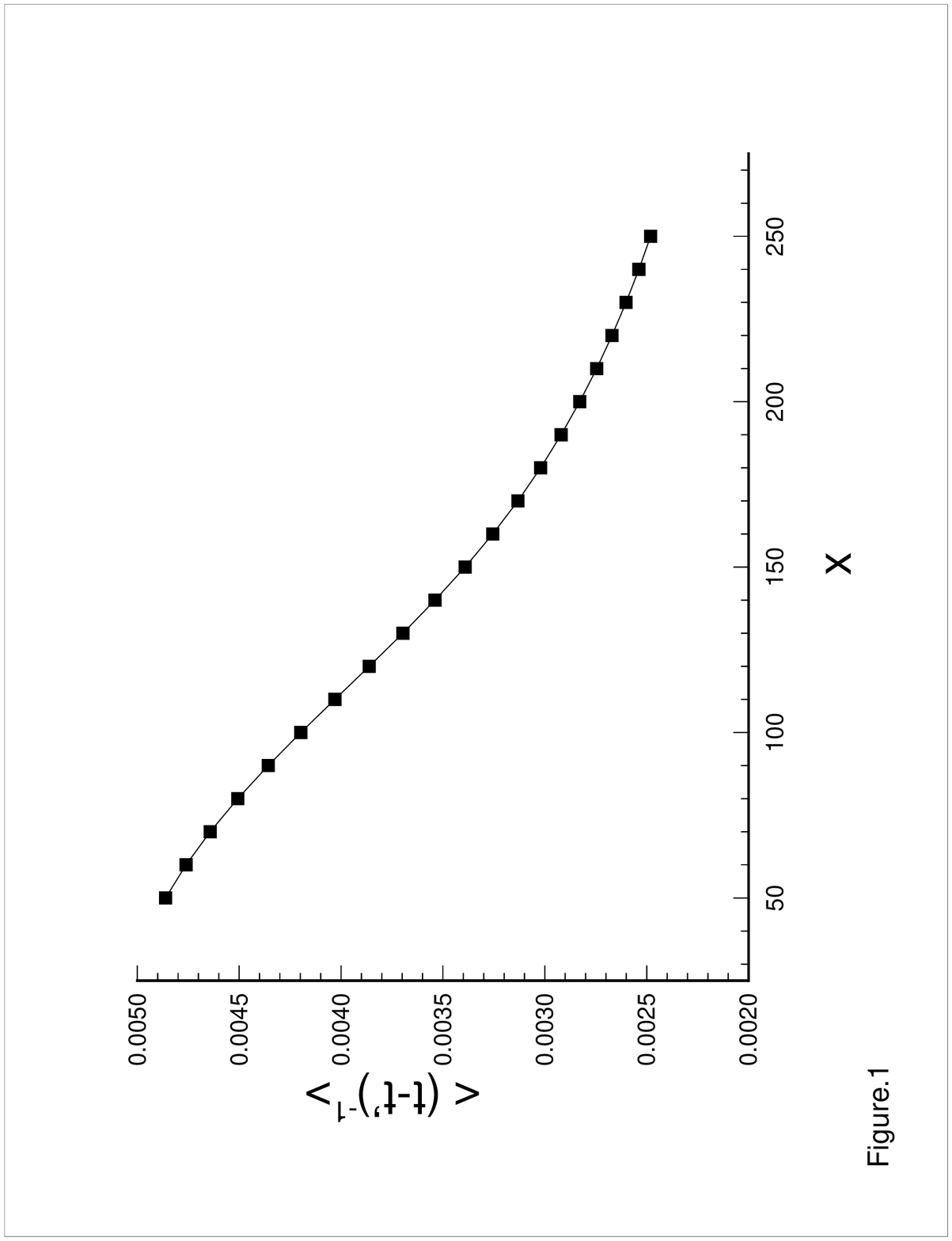}
\end{center}
\end{figure}
\begin{figure}
\begin{center}
\epsfxsize=10cm
\epsfbox{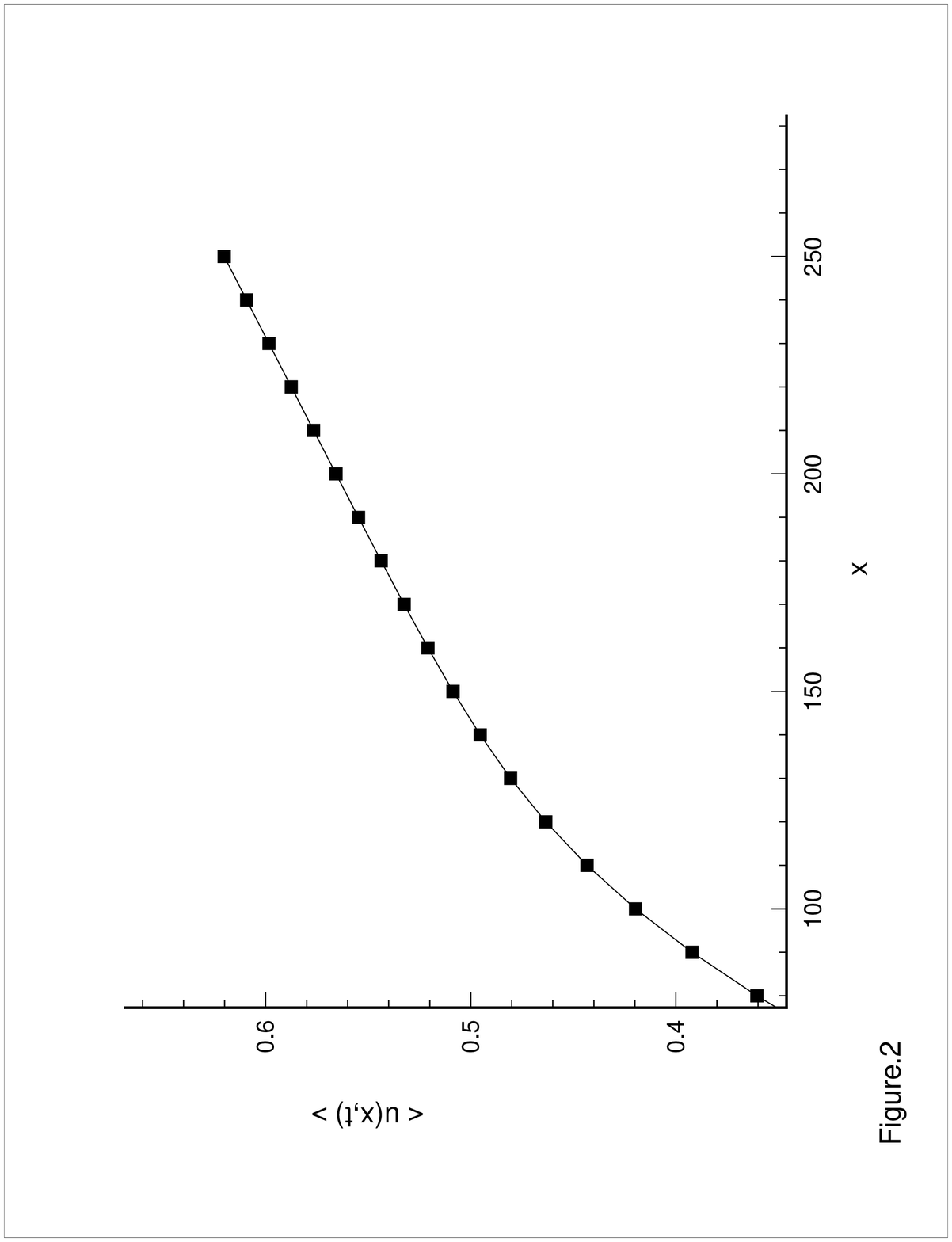}
\end{center}
\end{figure}

\end{document}